**Will the Mars Helicopter Induce Local Martian Atmospheric Breakdown?**

W. M. Farrell[1], J. L. McLain[1,2], J. R. Marshall[3] and A. Wang[4]
1. Solar System Exploration Division, NASA/Goddard Space Flight Center, Greenbelt, MD
2. Center for Research and Exploration in Space Sciences and Technology II, Univ. of Maryland, College Park, MD
3. SETI Institute, Mountain View, CA
4. Dept. Earth and Planetary Sciences & McDonnell Center for Space Sciences, Washington University in St. Louis, MO

**Abstract.** Any rotorcraft on Mars will fly in a low pressure and dusty environment. It is well known that helicopters on Earth become highly-charged due, in part, to triboelectric effects when flying in sandy conditions. We consider the possibility that the Mars Helicopter Scout (MHS), called Ingenuity, flying at Mars as part of the Mars2020 Perseverance mission, will also become charged due to grain-rotor triboelectric interactions. Given the low Martian atmospheric pressure of ~ 5 Torr, the tribocharge on the blade could become intense enough to stimulate gas breakdown near the surface of the rotorcraft. We modeled the grain-blade interaction as a line of current that forms along the blade edge in the region where grain-blade contacts are the greatest. This current then spreads throughout the entire connected quasi-conductive regions of the rotorcraft. Charge builds up on the craft and the dissipative pathway to remove charge is back into the atmosphere. We find that for blade tribocharging currents that form in an ambient atmospheric dust load, system current balance and charge dissipation can be accomplished via the nominal atmospheric conductive currents.  However, at takeoff and landing, the rotorcraft could be in a rotor-created particulate cloud, leading to local atmospheric electrical breakdown near the rotorcraft.  We especially note that the atmospheric currents in the breakdown are not large enough to create any hazard to Ingenuity itself, but Ingenuity operations can be considered a unique experiment that provides a test of the electrical properties of the Martian near-surface atmosphere.

I.      **Introduction**

The Perseverance rover will carry a first-ever rotorcraft designed to fly in the low pressure (~5 Torr) $CO_2$-rich Martian atmosphere. The rotorcraft, recently named Ingenuity, is approximately ~0.8-m in height and is lifted in the low-g environment by two ~1.2-m diameter blades revolving near 2800 rpm (Koning et al., 2019). It is well-known that rotorcraft in the terrestrial atmosphere develop substantial charge on their outer skin as they fly through the atmosphere. This charge develops, in part, due to tribo-electrification of the blade via collisions with aerosols and particulates in the atmosphere (Seibert, 1962; Pechacek et al., 1985; Zheng, 2013; Grosshans et al., 2018). Herein, we consider the possibility of charging of Ingenuity as it operates on Mars.

The low-pressure Martian atmosphere makes the gas highly susceptible to breakdown even under mildly active conditions. At pressures near 5 torr, atmospheric breakdown can occur at relatively low E-field values near 30 kV/m (compared to 3 MV/m at Earth) (Delory et al., 2006; Riousset et al., 2020). Given this low atmospheric breakdown field, there have been observational searches for evidence of impulsive filamentary discharges associated with Martian dust storms that are suspected to be electrically active due to mixing tribocharging dust and sand (Ruf et al., 2009; Gurnett et al., 2010, Anderson et al., 2011). However, recent laboratory studies suggests that the atmospheric dissipation of such charging systems may be in the form of an Townsend electron-avalanche opposed to an impulsive event (Farrell et al., 2015). Past laboratory studies also suggest that even mild dust mixing can give rise to a corona in a Mars-like environment. For example, Eden and Vonnegut (1972) put sand into a flask of $CO_2$ at 10 Torr and shook the flask creating glows and sparks in the gas. It has since been shown in the laboratory that mixing grains in a Mars environment will develop substantial charge (see recent report by Mendez-Harper et al. (2021) and references therein). Adding active human systems in the mixing process will also produces electricity. For example, Krauss et al. (2003) mixed sand on a low pressure (1-8 Torr) $CO_2$ gas using a stirring device, and detected discharges in excess of 10 per second at 1-2 Torr for simulated wind speeds of 2-3 m/s. If we consider the stirring device as an analog for a rotorcraft blade, then this experiment suggest that both the grains and the stirring device get electrified into a state of discharge.

As an analogy, it is well-known that terrestrial helicopters develop substantial charge in particulate-rich environments. Seibert (1962) reported on rotorcraft charging in blowing snow. Ice particulates charged the craft to many kilovolts. Similarly, very strong charging was reported for helicopters hovering over sandy terrain, with such rotorcraft developing potentials exceeding 100 kV (Pechacek et al., 1985). A model was developed (Pechacek et al., 1985) for the charge generated on a helicopter enveloped in a dense sand cloud. It was assumed that the grains in the cloud were tribocharged by grain-grain and grain-ground interactions. These pre-charged grains then transferred charge upon contact with the helicopter. Grosshans et al. (2018) simulated a helicopter blade interaction with a sand cloud, where the triboelectric charging occurred directly between the blade edge and the grains. They reported the development of triboelectric active regions along the blade edge where charge accumulated to levels of 0.8 µA/m for a blade tip slicing through the particulate-rich atmosphere at over 40 m/s.

Considering Mars atomsphere, besides being low pressure and thus susceptible to breakdown, the atmosphere also has an ambient dust load observed in non-dust storm season. Smith and Lemmon (2001) report an optical depth of ~0.5 for sunlight propagating through the atmosphere. For their assumed 13 km path length, this opacity corresponds to about 0.5 grains per cubic centimeter for 5-micron sized grains or 12 grains per cubic centimeter for 1-micron sized grains. We thus might anticipate that Ingenuity blades could also develop electrostatic charge by interacting with suspended particulates in the Martian atmosphere. Given the fast-rotating blades in this particulate-rich atmosphere, we would thus anticipate locations along the blades outer edge to develop local triboelectric charge due to the blade-grain interactions similar to that described by Grosshans et al. (2018) for terrestrial rotorcraft. Also, the rotating blade itself may intensify the local dust load around Ingenuity at take-off and landing which would further increase the charging effect, similar to the terrestrial rotorcraft case in sandy terrain described by Pechacek et al. (1985).

If the tribocharged rotorcraft develop large E-fields, then the local low-pressure Martian atmosphere could initiate a breakdown in the form of an electron avalanche and possibly even a persistent corona discharge (Farrell et al., 2015; 2017). In this work, we will describe a possible scenario of rotorcraft charging. We will derive an estimate for blade tribocharging currents, $J_T$, and associated rotorcraft charge that could drive an associated atmospheric (dissipation) current, $J_D$, possibly into a discharge regime.

## II. Breakdown of the Martian Atmosphere.

Laboratory measurements and modeling indicate that there are three regimes in a transition to atmospheric breakdown in a low-pressure $CO_2$ gas like that at Mars. **Figure 1** from Farrell et al. (2015) identifies these regions based on laboratory measurements of the breakdown of a 5 Torr $CO_2$ gas electrically-stressed by a uniform E-field having anode/cathode distances ranging from 0.8 to 6 cm. First, at low E-field strengths, the atmospheric dissipation currents, $J_D$, varies directly as $\sigma_o E$, where $\sigma_o$ is the nominal atmospheric conductivity ($\sim 10^{-12}$ S/m).

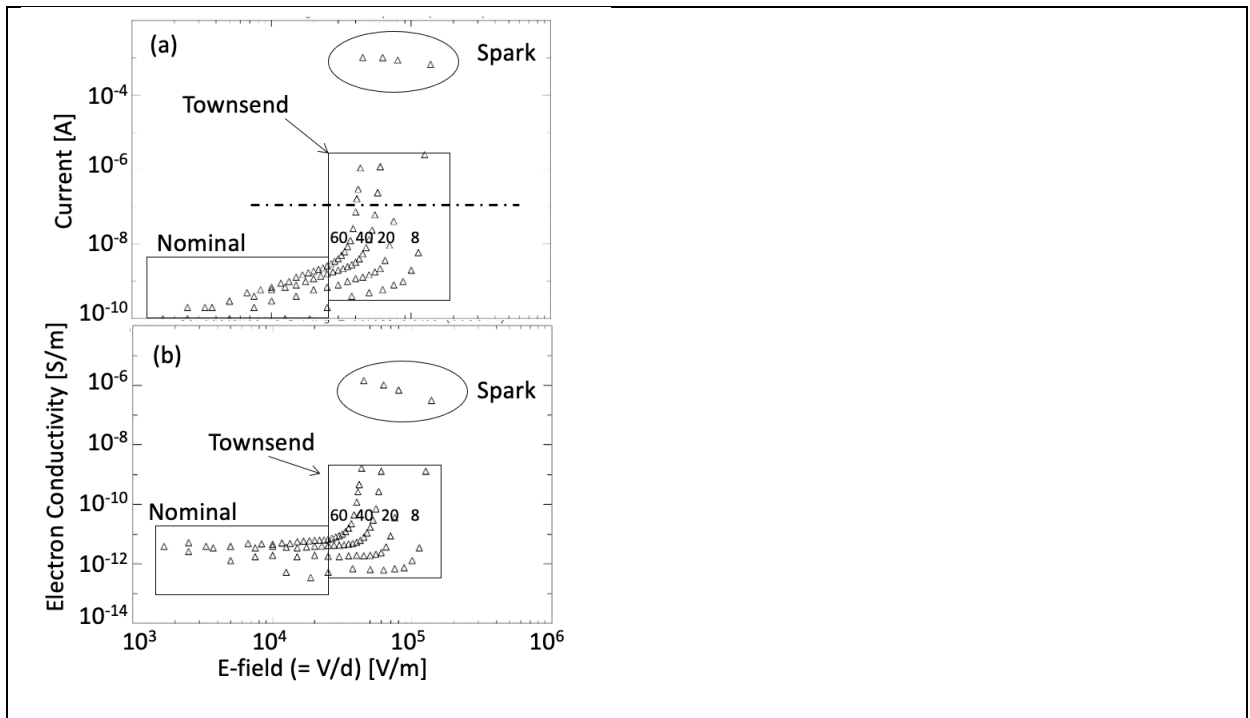

Figure 1. Lab-derived measurements of (a) induced current and (b) conductivity showing the stages of gas breakdown in a 5 Torr $CO_2$ atmosphere like that at Mars. Curves are for anode/cathode separations of 0.8- 6 cm. From Farrell et al. (2015).

Second, once the electric field becomes relatively large (> 20 kV/m), the free electrons in the gas undergo greater acceleration by the E field. In doing so, a population of electrons becomes energetic enough to initiate electron impact ionization upon collisions with the $CO_2$ gas molecules. These energetic electrons free up new electrons creating an electron avalanche. This regime is called the 'Townsend dark discharge' (Llewellyn-Jones, 1966). In this case, the conductivity exponentially grows as $\exp(\alpha d)$ where $\alpha$ is Townsend's primary ionization coefficient and d is the anode/cathode distance. The variable $\alpha$ represents the number of electron impact ionizations per unit length. The transition from nominal conductivity ($\sigma_o E$) to the Townsend regime is typically $\alpha d \sim 1$ or $J_D \sim \sigma_o e^1 E$. The variable, $\alpha$, exponentially increases with increasing E-field as (Llewellyn-Jones, 1966; Brown, 1966; Delory et al., 2006; Kok and Renno, 2009)

$$\alpha = \alpha_o \exp(-E_o/E) \qquad (1).$$

For a $CO_2$ gas at 5 Torr, laboratory analysis indicates that $\alpha_o \sim 10^4$/m and $E_o \sim 285$ kV/m (Table 1, Farrell et al., 2017).

Third, as indicated in Figure 1, once the E-field values become larger than about 50 kV/m, the conductivity will make an abrupt jump by a factor of 1000 to initiate a 'spark' discharge. The name 'spark' is a misnomer: The discharge is a continuous, visible, partially-ionized glow between the anode and cathode. As derived in Llewellyn-Jones (1966), the spark is triggered when additional secondary electrons also undergo an avalanche process, thereby providing a positive feedback that creates a current instability in the gas. In this case, the dissipation current has a denominator term

$$D = (1 - \gamma (\exp(\alpha d) - 1)) \qquad (2)$$

that incorporates the effect of the feedback process, where $\gamma = \omega/\alpha$ is the ratio of secondary-to-primary ionizations created in the gas and has a value of $\gamma = 0.01$-$0.02$ (Farrell et al., 2017). Note that the spark is triggered when $D = 0$ or, via Eq. (2), when $\alpha d = \ln(1/\gamma + 1) \sim \ln(1/\gamma)$ for $\gamma \ll 1$. For $\gamma = 0.02$, this corresponds to $\alpha d \sim 4$.

We thus can express a new effective conductivity along the E-field direction representative of the stressed gas near breakdown as (Farrell et al., 2015):

$$\sigma = \sigma_o \exp(\alpha d)/D \qquad (3)$$

This function is consistent with the profiles shown in Figure 1b. We can thus derive the atmospheric 'dissipation' current, $J_D$, that develops in the gas while under stress as:

$$J_D = \sigma_o \exp(\alpha d)\, E/D \qquad (4).$$

which is the trend shown in Figure 1a. We will use these laboratory results in Figure 1 and the associated model of atmosphere breakdown as a basis for understanding of the Ingenuity-atmosphere electrical interaction.

### III. Rotor Charging

We now consider the physical process by which a rotorcraft blade can get tribocharged in a dusty environment on Mars, paralleling the work of Grosshan et al. (2018) for blade tribocharging in a sandy terrestrial environment. In order to make this calculation, we make a set of assumptions that we further justify below: (1) There is an ambient dust load in the atmosphere and the rotor blade leading edge will be in contact with this dust. (2) At take-off and landing, there is a local pressure exerted upon the regolith to possibly create a particulate cloud that further enhances the dust load and the grain-blade edge interactions. (3) The grain-blade contacts undergo a tribocharge exchange (or contact electrification) with each grain passing charge to the blade edge. This grain-blade interaction creates a current that charges the blades and all surfaces conductively-coupled to the blades. As we describe in the next section, (4) the E-field associated with this tribocharge build-up increases until the local atmosphere undergoes breakdown. At that time, the return currents from the atmosphere attempt to come into balance with the tribocharing currents, creating an equilibrium E-field. We consider a case where the tribocharging current is constant and a second case where the tribocharging current increases in time -both cases are presented to demonstrate the atmosphere's attempt to bring the system to current equilibrium.

**Figure 2** illustrates the charging of an Ingenuity blade moving through the dusty gas. Like Grosshan et al. (2018), we assume the grain-blade collisions occur at the wind-facing blade edge. Specifically, as the blade moves through the gas, area $\zeta$ along the blade edge charges due to triboelectric contacts between the grains and blade edge. This charging is a function of the rotor velocity, and thus the entire outer edge of the blade charges – forming a line current defined by the blade edge. However, the highest levels of charging are expected at the tip, or end of the line charge, since that region is moving at the greatest velocity though the dusty gas and thus has the highest incident dust flux. Figure 4 of Grosshans et al. (2018) also shows that the region near the end of the blade is the most triboelectric active location, with charging varying directly as a function of position along the length of the blade. For Ingenuity, since there are 4 blades, all 4 will become charged.

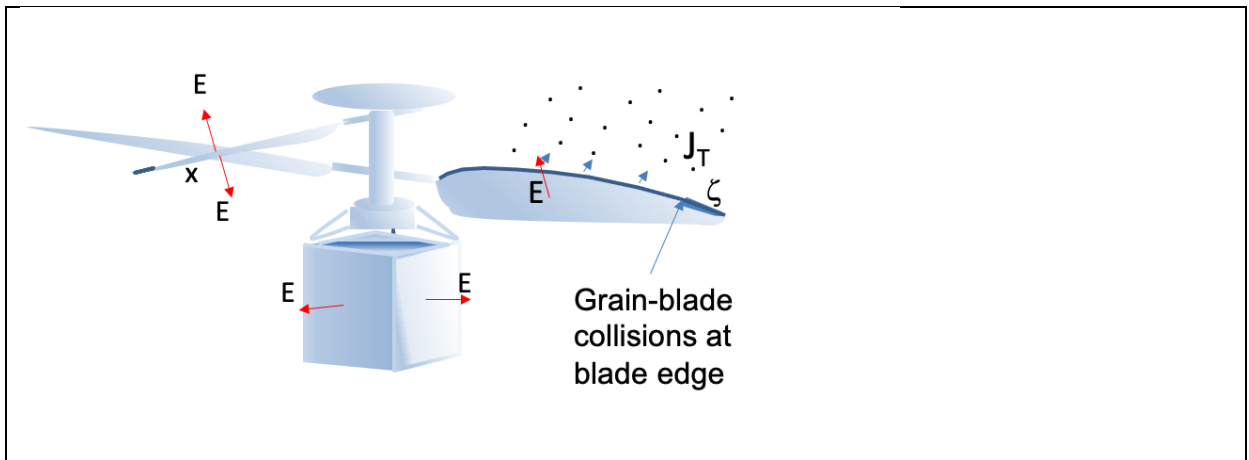

Figure 2- An illustration of the blade configuration for Ingenuity. The blade edge incident with the dust grains forms a line of current. This current then spreads throughout the blade and other conductively-coupled surfaces, giving rise to a normal-directed E-field.

Grosshans et al. (2018) considered the charge build-up on the blade edge and did not consider the composition of the blade and possible redistribution of charge to other parts of the rotorcraft via current flow on conductive surfaces. Given that the Ingenuity's blade is a carbon composite material, which is generally mildly conductive, we consider that the tribocharge at the blade edge quickly spreads, via conduction, along the blade surface and to all connected conductive surfaces of the rotorcraft itself. Figure 2 illustrates the normal-directed E-field vector from the flatter surface of the rotorcraft that should develop as the tribo-charge is redistributed across the system. We will examine the evolution of this E-field in the case where the blade has a conductive pathway to other conductive surfaces on the rotorcraft (and also for the case where the blade is electrically isolated from the body).

As suggested by Figure 1, once the charge on these surfaces builds up to a point that the E-field near the surface is on the order of ~30 kV/m, the stressed atmosphere should initiate a breakdown.

It should be noted that the E-field configuration illustrated in Figure 2 has both an equilibrium (DC) and AC component. The blades of the two rotors counter-rotate such that blades from the top rotor pass in proximity of a bottom blade twice per revolution. Since the blades are of the same charge polarity, this will result in a periodically-occurring perturbation of electric flux between the blades. This local inter-blade E-field perturbation occurs at twice the blade revolution (since there are two blades) creating a ~ 4800/minute (or ~80 Hz) fluctuation in the local inter-blade E-field. We first quantify the quasi-DC E-field that develops by charge generation, and then consider this 80 Hz blade-blade interaction as a perturbation E-field overlaying on the equilibrium value.

Regarding the quasi-DC aspects of the E-field formation: Consider, first, a finite small area along the blade edge out near tip of the blade having area, $\zeta$. The tip of the blade where this small area resides is located a distance $z \sim 60$ cm from the rotor shaft (see Figure 2). This area of the blade edge will slice though the dusty Mars atmospheric gas at velocity, $v = r\omega$, with $\omega$ being the revolution rate of about $\omega \sim 290$ revolutions/s (for 2800 RPM). The blade tip is traveling at about 177 m/s through the gas at a Mach number of 0.76 (Koning et al., 2019). The dust flux onto surface area $\zeta$ along the blade edge is then

$$F = n_d\, v = n_d\, r\, \omega \qquad (5)$$

where $n_d$ is the dust density (for ambient atmospheric dust load $\sim 10/cm^3$, see introduction section) and r is the distance along the blade. The tribocharging current (i.e, charging influx) incident on the blade is thus

$$J_T = q_T\, n_d\, r\, \omega \qquad (6)$$

where $q_T$ is the tribocharge transferred at each grain/blade contact. This charging occurs along a thin region along the blade edge and the current generation increases in current density with increasing distance, r, along the blade (Grosshans et al., 2018). In essence, the blade edge is a current source that acts to charge the other conductive regions of Ingenuity.

Equation (14) of Desch and Cuzzi (2000) provides an estimate of the tribocharge transfer between two grains, $q_T$. We can modify this equation assuming that the reduced radius between a very large structure and a small grain is effectively the small micron-sized grain. In this case, the tribocharge transfer scaling equation in Desch and Cuzzi applied to a 1 micron sized grain reduces to:

$$q_T \sim (\Delta\Phi/2V) \text{ fC} \qquad (7)$$

where $\Delta\Phi = \Phi_G - \Phi_B$ is the difference in triboelectric surface potential between the grain and the blade. In general, the surface with the larger triboelectric potential will tend to charge negative. Dielectrics tend to have values below $\Phi < 3$ V while metals have values near $\Phi \sim 4$ V. Hence, for large compositional difference (large work function difference) between the blade and grain, we can expect $\Delta\Phi \sim 1V$, allowing the blade to collect $q_T \sim 0.5$ fC per grain contact, assuming a micron-sized grain. The value of $q_T$ varies directly with grain radius (Desch and Cuzzi, 2000), thus larger grains will generate a proportionally larger contact charge. The charge polarity of the blade tip is a function of the composition of the blade and grains (their difference in electronic work functions). Pechacek et al. (1985) reported that terrestrial rotorcraft charged to large positive values when interacting with sand clouds, and we thus assume a similar positive blade charge herein. We note that Grosshans et al. (2018) found on average close to 200 pC of charge being transferred per blade-grain collision. However, they were modeling a distribution of grains with a mean diameter of 800 microns – substantially larger than those modeled herein. In the Discussion and Conclusions section, we will further describe the expected uncertainty in the rotorcraft charging process based on different grain-blade tribocharge transfer assumptions.

Given the above, we can estimate that the triboelectric charging current density at the tip as $J_T$ (r = 0.6 m) $\sim 0.9$ $\mu A/m^2$ for an ambient atmospheric micron-sized dust load of $n_d \sim 10^7/cm^3$. $J_T$ will decrease along the blade edge linearly with decreasing r. However, during take-off and landing, the rotor's winds could lift grains off the surface and temporarily add to the local dust load. For example, if the dust lifted off the surface is closer to $100/cm^3$ (similar to the micron-sized dust load in dust devils (Farrell et al., 2003)), then the blade slicing through this enveloping dust cloud will increase $J_T$ near the tip to $\sim 10$ $\mu A/m^2$. We provide further justification of a possible rotor-driven particulate cloud in the Appendix 3.

To get the total current generated at each blade edge, $I_{blade}$, we performed a surface integral using the current density defined Equation (6) and found

$$I_{blade} = \tfrac{1}{2}\, q_T\, n_d\, \omega\, \Delta x\, L^2 \qquad (8)$$

with $\Delta x$ being the thickness of the grain-impacting edge of the blade (~0.01m) and L being the length of the blade (~0.6 m). This total current is about 2.7 nA for an ambient atmospheric dust load, and possibly as large as 27 nA for any enhanced dust loads at take-off and landing.

Hypothetically, if this charging electrical system were placed in a gas with no intrinsic conductivity, there would be no electrical dissipation loss into the atmosphere. In this case, the surface charge, $\eta$, on all connected conductive elements of Ingenuity would continually grow as

$$\Delta \eta = 4\, I_{blade}\, \Delta t / A_{MHS} \qquad (9)$$

where $A_{MHS}$ is the conducting area of the rotorcraft that is electrically connected to the 4 current-generating blade edges. The electric field in close vicinity to flat conducting surfaces becomes

$$E \sim \Delta \eta / \varepsilon_o \sim 4\, I_{blade}\, \Delta t / (\varepsilon_o A_{MHS}) \qquad (10).$$

For $I_{blade}$ = 2.7 nA and an exposed conducting area of $A_{MHS} \sim 1\ m^2$ (see Appendix 1), the E-field close to flat surfaces will grow as ~ 1200 $\Delta t$ V/m and within 30 seconds of operation, the E-field at flat surfaces will be near 36 kV/m.

### IV.  Blade Dissipation through the Atmosphere

While the derivation above applies for a non-conductive medium, in reality, Ingenuity will operate in a conductive gas and thus there is charge dissipative loss back into the atmosphere. Further, as illustrated in Figure 1 and Equations (3-4), the conductivity and dissipation currents at Mars are a complex function of the E-field itself. Hence, the dissipation current to remove charge from Ingenuity, $J_D$, described by Eq. (4) is a function of the E-field generated by tribo-charged induced charge density, $\Delta \eta$, which itself is a function of tribocharging current onto the blade, $J_T$: $J_D = J_D(E(\Delta \eta(J_T)))$. We can write a continuity equation to describe the balance of the tribo-charging induced surface charge and dissipation currents on a flat surface of Ingenuity as:

$$\varepsilon_o dE/dt = \Delta\eta/\Delta t - J_D \qquad (11)$$

with dissipation current density as $J_D = \sigma_o \exp(\alpha d) E/D$ (from Eq. (4)). For a $\Delta\eta/\Delta t$ that is constant in time (i.e., second derivative of the charge density is constant and neglecting the 80 Hz signal), an equilibrium E-field, $E_{eq}$, can be found such that Eq. (11) is zero and $\Delta\eta/\Delta t \sim J_D$. This current balance implies that the atmosphere is removing charge from Ingenuity's exposed conductive surfaces in direct proportion with the blade tribo-charge build-up. In equilibrium, $J_D = \varepsilon_o E_{eq}/\tau$, where $\tau$ is the dissipation time to reach the equilibrium and $\tau$ is derived as well.

**Figure 3** shows the equilibrium E-field solving for $\Delta\eta/\Delta t = J_D$ from a flat surface on Ingenuity (e.g., on the blade face or flat body surface in Figure 2) as a function of the tribo-electrically generated surface charging rate, $\Delta\eta/\Delta t$. For increasing blade surface charging rate, the equilibrium E-field increase linearly up to a value of $\Delta\eta/\Delta t = 3 \times 10^{-8}$ A/m$^2$. At this point, surface charge on the flat surface creates an E-field that is large enough to induce a Townsend dark discharge. Above $\Delta\eta/\Delta t = 3 \times 10^{-8}$ A/m$^2$, the conductivity exponentially increases (e.g., Figure 1) according to $\sigma = \sigma_o \exp(\alpha d)$ with $\sigma_o = 10^{-12}$ S/m, with $\alpha$ described by Eq. (1), and d set to 2 cm above the flat surface. As illustrated in Figure 3, above $\Delta\eta/\Delta t = 3 \times 10^{-8}$ A/m$^2$, the E-field no longer displays a linear relation with $J_C$, and instead varies slowly from 40 kV/m to ~72 kV/m over four decades of $\Delta\eta/\Delta t$. The values remain quasi-constant even as the blade charging rate $\Delta\eta/\Delta t$ continues to exponentially increase. The reason for the limiting effect: In equilibrium, the breakdown currents formed in the Townsend dark discharge are capable of offsetting any excess tribocharge collected by the blade thus keeping E and total $\eta$ quasi-constant. The total surface charge ($\eta_{total} \sim \varepsilon_o E$) in any exposed flat area is then the combination of the redistributed tribocharge ($\Delta\eta/\Delta t$) along with the negating charge drawn from the atmosphere. Note that above $3 \times 10^{-8}$ A/m$^2$, the near-surface electron density and conductivity both increase exponentially with increasing $\Delta\eta/\Delta t$, thus the added electrons in the electron avalanche offset the charge created by the grain-blade interaction. In the Townsend regime, the equilibrium time scale decreases exponentially reflecting the fact that the added electron population can quickly bring the charging system into equilibrium. As the blade current continues to increase, D drops below unity and the likelihood of a spark or corona-like discharge increases – which would be a visible manifestation of the breakdown. Recall that the spark or corona condition is met as D approaches 0 (see Equation 2).

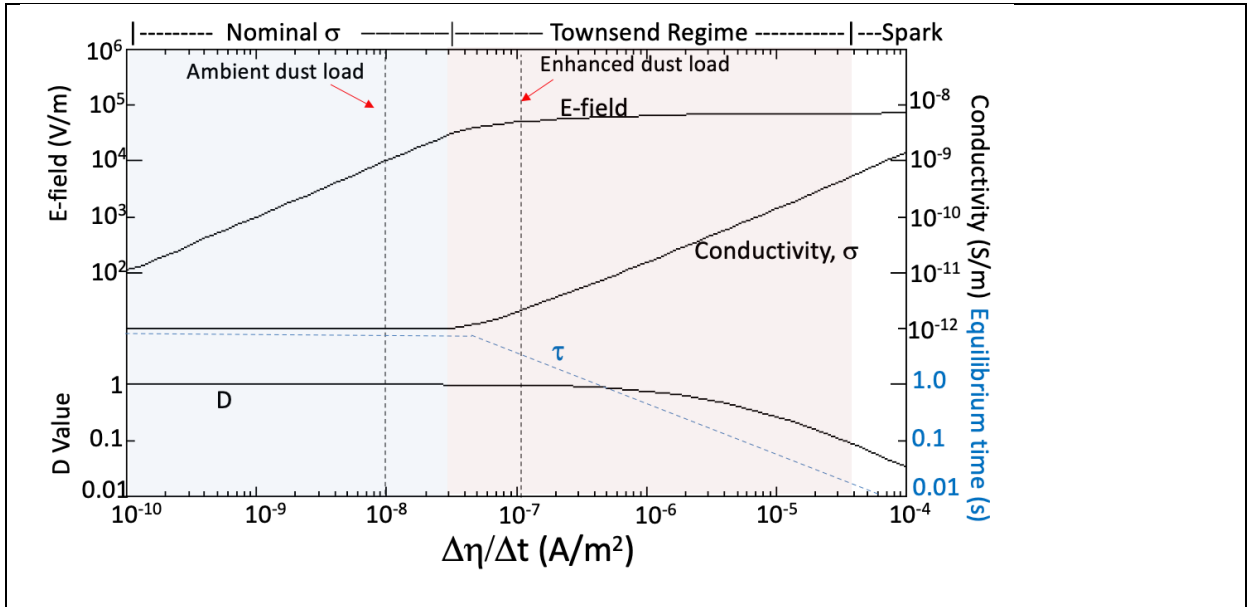

Figure 3- The equilibrium E-field, associated atmospheric conductivity, equilibrium time, and D value as a function of the conductive surface charging current created by grain-blade interactions.

We also indicate in Figure 3 the expected value of $\Delta\eta/\Delta t$ under ambient ($10/cm^3$) and possible enhanced ($100/cm^3$) take-off and landing atmospheric dust load conditions. Note that for typical dust load conditions, an equilibrium between charging and dissipation currents is reached near an E-field of 10 kV/m. In this case, the atmospheric conductivity is $10^{-12}$ S/m and not enhanced above nominal values. An electron avalanche and Townsend discharge are not initiated. In equilibrium, the atmosphere possesses enough intrinsic charge to offset the blade charging, albeit a relatively large E-field is needed to draw in enough atmospheric charge to bringing the two currents into equilibrium.

However, if a dust or sand cloud surrounds Ingenuity at take-off and landing, greater charge could be generated from blade-grain interactions. In Figure 3, we show a case for this enhanced dust load where the dust density increases by a factor of 10 (to $100/cm^3$). In this enhanced case, the E-field is near 50 kV/m. The atmospheric conductivity is predicted to be 2-3 times larger than nominal values due to the presence of an enhanced electron population and the formation of an electron avalanche. The system is now dissipating via the currents formed in a Townsend dark discharge. The extra electrons in the discharge are needed to offset the enhanced tribocharging. The initiation of breakdown occurs in this case. We thus anticipate higher electron densities, along with higher $CO_2^+$ densities due to impact ionization. There should also be higher $CO^-$ concentrations due an increase in electron dissociative attachment (Delory et al., 2006; Jackson et al., 2010).

We can also consider a second charging scenario where the blade is not electrically connected to the rest of Ingenuity, such that the blade alone charges up in isolation. In this case, the conducive currents passing over the blade's surface is $\Delta\eta/\Delta t = I_{blade}/A_{blade}$, with the area of the blade at approximately ~0.12 m$^2$. While the area is reduced by a factor of ~ 8 compared to the assumed larger Ingenuity conducting area, the isolated blade will not obtain any current created from the other three blades. Thus, the total surface charging rate along the blade surface used to formulate Figure 3, $\Delta\eta/\Delta t$, drops by a factor of 4. The combined effect is to shift the ambient and enhanced dust load lines in Figure 3 to the right by a factor of 2. In this single blade-only case, the atmospheric response to tribocharging for the ambient dust load is still in the nominal conductivity regime and the enhanced load case moves further into the Townsend regime.

Given the equilibrium E-field as a function of $\Delta\eta/\Delta t$ shown in Figure 3, we can now address the effect of the 80 Hz fluctuations on charging/discharging from blade-blade interactions. This perturbation analysis is similar to examining the AC signal fluctuations (small signal analysis) for a diode or transistor, where the equilibrium levels are established first and then AC fluctuations/small signals about equilibrium are quantified using current-versus-voltage curves. We can consider the passing of the two blades, an upper and lower blade (see Figure 2), as two charged disks passing each other. The electric field in the close vicinity of one of the blades is then $E = E_{eq} + \delta E$, where $\delta E$ is the perturbation in the electric field from the other passing blade.

Consider the E-field in proximity to the bottom side of the upper blade (position 'x' in Figure 2). In close proximity to this surface, the E-field will have a value $E_{eq} \sim -\eta_{upper}(\varepsilon_o)^{-1}$, where $\eta_{upper}$ is the surface charge on the upper blade with the minus sign indicating that the E-field is pointing downward (nadir). However, at position 'x', there will be a perturbation as the lower blade passes, with an E-field from the top-side of this lower blade at its maximum value being $\delta E \sim +\eta_{lower}(\varepsilon_o)^{-1} w/\pi z$ where w is the blade width and z is the blade-blade separation (see Appendix 2). The plus sign indicates that the E-field is pointing upward (zenith), in an opposite direction from the E-field along the bottom-side of the upper blade. For a separation of the two blades at z ~ 0.1 m, w ~ 0.05 m and $\eta_{upper} \sim \eta_{lower}$, this E-field from the lower blade represents a perturbation of about ~20% of the E-field in proximity to the upper blade. Near position 'x' in Figure 2, the E-field perturbation from the lower blade will act to temporarily decrease E by up to 20%.

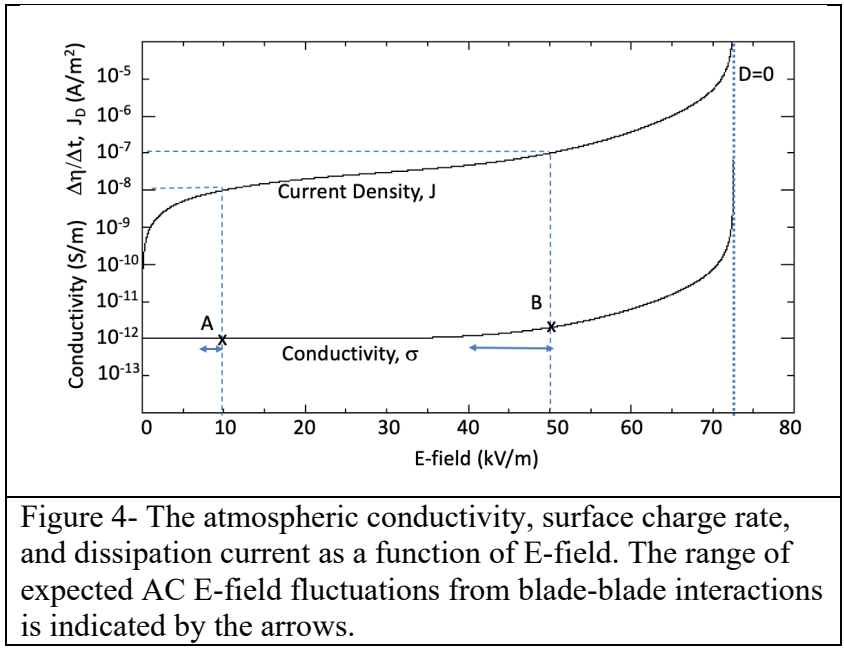

Figure 4- The atmospheric conductivity, surface charge rate, and dissipation current as a function of E-field. The range of expected AC E-field fluctuations from blade-blade interactions is indicated by the arrows.

**Figure 4** shows the atmospheric conductivity and surface charging rate (and dissipation currents, since $\Delta\eta/\Delta t = J_D$ in equilibrium) for a flat surface as a function of driving E-field for the Mars atmosphere with $\sigma_o = 10^{-12}$, d = 0.02 m and $\gamma = 0.02$ (similar to the conditions used in Figure 3). In essence, this figure is a model result that is representative of the lab data shown in Figure 1.

Consider the equilibrium point at A. This location is driven by a surface charging rate of $10^{-8}$ A/m² resulting from blade interactions with an ambient dust load. The equilibrium E-field is near 10 kV/m. The arrow represents a -20% perturbation near one of the blades resulting from the blade-blade interactions. Note that with the passing of the perturbing blade, the atmospheric conductivity remains at nominal values.

Point B represents the system driven by a surface charging rate of $10^{-7}$ A/m² like that modeled for a possible enhanced dust load at take-off and landing. The equilibrium E-field is near 50 kV/m. Note that the atmospheric conductivity is in the Townsend regime and there is a breakdown in the form of an electron avalanche. The arrow below B represents a perturbation E-field from blade-blade interactions that is -20% of the equilibrium value. Even with the perturbation, the discharge remains in the Townsend regime, although the breakdown is reduced with the passing blade.

We conclude that breakdown is most likely when the co-rotating blades are perpendicular to each other (i.e., 90º apart). The effect of the passing blade (when 0º apart) is to suppress breakdown by reducing the E-field in the region between the blades.

## V.     Discussion and Conclusions

Regarding syntax: For D= 0 in a uniform field between two plates, like that in the laboratory work shown Figure 1, the visible discharge is referred to as a 'spark' discharge. However, this same D=0 condition induced in a nonuniform E-field geometry, usually in the gas about a small, highly charge conductor like a rod or line charge, is referred to as a corona (Llewellyn-Jones, 1966). Herein, we have performed E-field calculations assuming the sensing point is in relatively close proximity to flat surfaces. However, the E fields are really non-uniform with distance due to the finite size of the surfaces themselves. This non-uniformity of the E-field also occurs at edges of surfaces and along thin cylindrical pieces. Thus, any stimulated 'glow' from Ingenuity is considered a corona.

We solved Eq. (11) assuming the surface charging rate, $\Delta\eta/\Delta t$, is constant (i.e., the second derivative in time of the surface charge is zero). However, if this rate is itself varying in time, the evolution of the E-field has to be described by the time derivative of Eq. (11) which can be written as:

$$\varepsilon_o \frac{d^2E}{dt^2} + \sigma \frac{dE}{dt} + E \frac{d\sigma}{dt} = \frac{d^2\eta}{dt^2} \qquad (12)$$

where $\sigma$ is defined by Eq. (3). Generally speaking, for low pre-breakdown E-field values, $d\sigma/dt$ is small since $\sigma$ has a uniform value of $10^{-12}$ S/m below the steep $d\sigma/dE$ region in Figures 1 and 4. Thus the third term on the left side of the equation is small for modest E-field values. If also the temporal variation of the surface charging rate, $\delta t$, is faster than the nominal atmospheric dissipation time, $\varepsilon_o/\sigma$, then the second term on the left is small. The result is the trivial solution $\varepsilon_o$ dE/dt ~ d$\eta$/dt. In this case, the E-field will grow directly with the growth of the surface charge (e.g., like Eq. (10)).

However, this fast-growing E-field solution will not continue to infinitely large values. Instead, it will continue until the E-field reaches a point where breakdown occurs. At this breakdown point, the d$\sigma$/dt factor in the third term of Eq. (12), which itself is a function of the E-field, becomes large and thus acts to shunt the growth of the E-field. Integrating Eq. (12) once we obtain:

$$\varepsilon_o \frac{dE}{dt}(t) = \frac{\eta_o}{\tau} e^{t/\tau} - \int_0^{t-dt} \sigma(t) \frac{dE}{dt}(t) \, dt - \int_0^{t-dt} \frac{d\sigma}{dE}(t) E(t) \frac{dE}{dt}(t) dt \qquad (13)$$

where we consider an exponentially-growing charging current, $\eta = \eta_o \exp(t/\tau)$, with $\tau$ being the e-folding growth time of the current, and where we have re-written d$\sigma$/dt as (d$\sigma$/dE) (dE/dt) in the third term on the right. The limits of integration extend from t = 0 to t = t - dt, indicating that we applied the preceding solution at t - dt (i.e., E(t-dt) and dE/dt(t-dt)), to estimate dE/dt at time t. The first term on the right of Eq (13) represents the E-field contribution from the exponentially-growing surface charge flowing through flat surface (for example, the blade face) while the second and third terms represent the compensating effects of current drawn from the atmosphere.

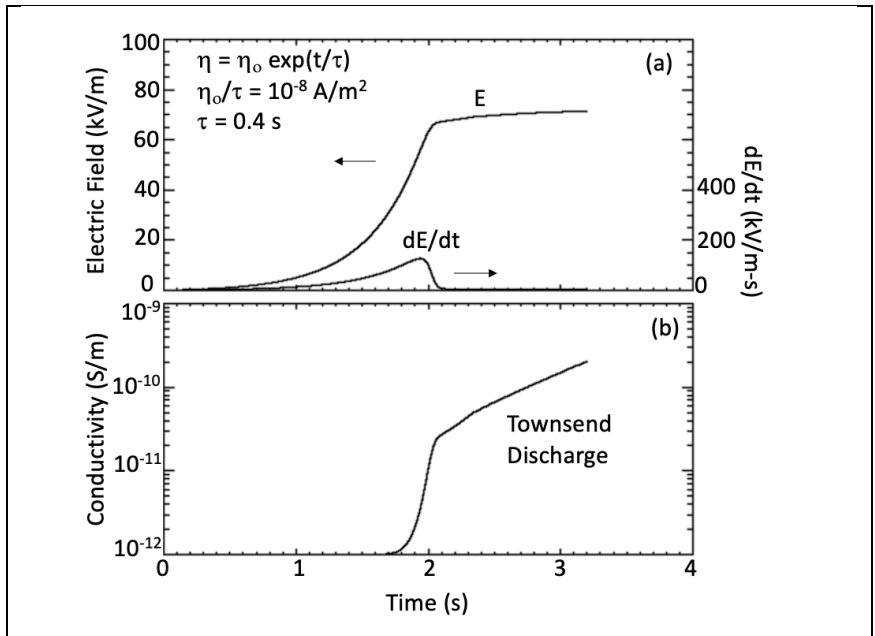

Figure 5- The (a) growth of the E-field and its time derivative for an exponentially-growing charging current (solving Eq. (13)) along with (b) the associated increase in near-surface electrical conductivity.

Consider a case where the rotor is initially turned on. During this time period we might assume the grain impacts from both ambient dust and rotor-lofted grains are exponentially-increasing in incidence and that both E and dE/dt are zero at the start. **Figure 5** shows a solution for Eq. (13) for σ defined by Eq. (3) with $\sigma_o = 10^{-12}$, d = 0.02 m and γ = 0.02 and with the tribocharging of the system having an e-folding growth of τ = 0.4 s. We assume an initial charging current in the conductors as $\eta_o/\tau = 10^{-8}$ A/m² (i.e., the initial charging current is set at the ambient dust load, and then it is allowed to exponentially increases in time thereafter).

As illustrated in the solution shown in Figure 5a, from t=0 to 1.7 seconds, the E-field and dE/dt both grow exponentially in time directly with the exponential growth of the surface charging rate, dη/dt. In this period, when considering Eq. (13), the last two terms on the right-had side are small compared to the first term (i.e., the current growth term). However, when the E-field reaches a value near 70 kV/m, the last term on the right in Eq (13) becomes large and comparable to the growth term, effectively shunting the growth of the E-field. Near t= 1.9 s, dE/dt drops to low values and the E-field values become quasi-constant. In Figure 5b, we see that the near-surface conductivity jumps by a factor of 30 from nominal values at $10^{-12}$ S/m to 3 x $10^{-13}$ S/m between 1.7 and 2 seconds. Thereafter, the conductivity exponentially-rises in cadence with dη/dt; this occurring to keep the first and third terms of Eq (13) near equal values and dE/dt relatively small.

While Figure 5 illustrates one non-unique example, the results imply generally that at take-off and landing, when the dust density is quickly increasing due to particulates being lifted by the rotors, there could be a fast temporal change in the surface current rate - a substantial $d^2\eta/dt^2$ - that triggers a Townsend discharge or even a corona discharge an in attempt to reduce the growth of the E-field. Thus, a corona may reveal itself during this take-off and landing period when there is a substantial $d^2\eta/dt^2$.

Pechacek et al. (1985) considered a different process for helicopter charging over sandy surfaces: That the mixing grains in the rotorcraft winds are themselves tribocharged by grain-surface and grain-grain interactions. The lifted charged grains then enshroud the helicopter in a charged grain cloud which transfers charge to the helicopter upon grain-helicopter interaction. They did not consider the possibility of direct tribocharging from the grains to the moving blades on the rotorcraft. We did not address this alternate charging possibility here – instead addressing the charging of the blade tip by the direct grain-blade tribocharging effect (similar to Grosshans et al. (2018)). We thus note that our scenario for charging may not be analytically unique – there may be other ways to charge Ingenuity in the low-pressure dusty atmosphere, especially when enshrouded in a dust cloud. Herein, we simply follow one reasonable possibility to its conclusions.

There may be more vigorous charging mechanisms than grain-blade tribocharging. The blades may be rotating fast enough to create physical damage to the grains and blades (i.e., chip edges, etc), and thus charge via grain destruction rather than the milder contact electrification/tribocharging. For example, very little scientific literature currently exists on the physics of blade charging by the Kopp-Etchells Effect where dust grains impact helicopter blade edges with enough energy to erode the blade and create sparks, thus forming a glow in a dusty environment that intensifies upon take-off and landing (Collins and Moore, 2014). Such grain-destructing events could potentially generate more charge per grain than the tribo-electrification processes applied here.

We note that there are a number of tribocharging models in the literature. We apply that of Desch and Cuzzi (2000) since it includes effects from tribocharging that incorporate both size and compositional differences. Melnik and Parrot (1998) assumed a charge transfer between two grains of approximately 1 fC per micron radius of the grain, which is comparable to the charge transfer applied herein. For contacts between two insulators, an advanced model by Kok and Lacks (2009) has been developed. However, in our case herein, the blades are conducting carbon composite interacting with insulating grains. There are also laboratory studies of grain-grain tribocharging that derive the charge per grain (for example, Krauss et al. (2006) and Mendez-Harper et al. (2021)). Differences between all of these studies represent an inherent uncertainty in the efficiency of the tribocharging process which then creates an uncertainty in our results. Fortunately, Figure 3 provides a way to adjust the result to differences in charging from the applied Desch and Cuzzi model. For example, if the charge transfer process is more vigerous by a factor of 100 as compared to the Desch and Cuzzi formalism, the $\Delta\eta/\Delta t$ value in Figure 3 for the nominal case would shift to the right by a factor of 100 into the Townsend regime. If the charge transfer process is less efficient by a factor of 10 as compared to the Desch and Cuzzi model, the $\Delta\eta/\Delta t$ value in Figure 3 for the nominal case would shift to the left by a factor of 10. Hence, Figure 3 allows one to examine the system-level changes for a tribocharge exchange that has a different charging efficiency than the one applied herein.

Despite the inherent uncertainty in the tribocharging process, the Ingenuity system provides a unique opportunity to examine the Mars atmosphere under an electrically stressed situation. One can think of Ingenuity as a charge object – like a Van de Graaff generator – that could induce local breakdown of the gas. This object will allow a testing of previous modeling and laboratory work. We note that an ideal method is to fly an atmospheric electricity package to the Martian surface to measure the E-fields and breakdown thresholds in wind-blown dust features. However, this rotorcraft experiment allows an indirect method to examine the electrical breakdown processes in the atmosphere. We note that any human or natural system that creates mixing dust should find relatively large E-fields in the active region. Electrometers and RF discharge-sensing devices should thus be included on future payloads to search for manifestations of breakdown. For example, rover wheels moving over the Martian soil should charge up, creating a local E-field that might induce a corona or even a mini-discharge (like that described in Krauss et al. (2003)). Thus, a small electrometer plate and small RF antenna could be placed near the wheel to measure the E-field and associated discharge. We thus suggest that there are ways to examine the atmospheric electricity at Mars with small, inexpensive sensors placed near systems that interact with the soil/regolith that should be exploited on future missions. They may not provide the full picture, but can at least confirm the manifestation of atmospheric breakdown via in situ sensing.

In summary, our laboratory breakdown studies of a Mars-like low pressure $CO_2$ gas (Figure 1) are applied to the case of charging of the Ingenuity rotorcraft while operating on Mars. We find that breakdown is possible especially during take-off and landing when the dust load in the atmosphere about Ingenuity could be higher than normal. We suggest that operations occur once at twilight so that any corona-like glow be observed by Perseverance in the low light environment. Perseverance's imaging systems should be monitoring Ingenuity especially at takeoff and landing to search for a possible corona induced on the blades in the low-pressure Martian atmosphere that might confirm the atmospheric electricity laboratory experiments like that in Figure 1. We make a special note that the triboelectric current levels being described herein will not damage Ingenuity itself. However, Ingenuity provides a unique opportunity to be used as a scientific experiment unto itself to examine the conditions for Mars atmospheric breakdown. At no other time in history have we ever had such a highly-active human system stirring and mixing the dusty, low pressure Martian atmosphere.

**Appendix 1: Approximate Surface Area of Ingenuity**

We estimated that the conducting surface area of Ingenuity is $A_{MHS} \sim 1$ m$^2$ based on photographs and drawings. The 4 blades together have 8 sides (top and bottom) at approximately 62 cm in length and at their widest near ~ 12 cm, corresponding to a total area of about 0.6 m$^2$ (see Figure 4 of Pipenberg et al., 2019). The main body is a cube of approximately 14-cm in height, width, and length giving a surface area of 0.12 m$^2$. The solar panel area is approximately a 20 cm x 10 cm surface and for two exposed sides is approximately 0.04 m$^2$ – although these are likely not conductive. The rotor shaft is very roughly 20 cm in length and has a radius of approximately 3 cm (although it varies along the shaft) giving an area of 5 x 10$^{-4}$ m$^2$. The area of the 4 thin legs and spherical-like landing pad are less than 0.03 m$^2$ in total. The total area via this very rough analysis is slightly less than 0.8 m$^2$, and we assume there are other smaller appendages and exposed regions that justify the use of $A_{MHS} \sim 1$ m$^2$. We assume that these regions are conductive and all electrically-connected from the blade to the legs. In the text, we also consider the situation if the blade is electrically-isolated from the rest of the rotorcraft.

**Appendix 2: E-field from a long, thin conductor with surface charge, h**

We solved for the vertical E-field in the central region of a rectangular conductor that is representative of the blade. The length of the conductor, L, is assumed to be much larger than the conductor width, w, and the blade-blade distance, z: L >> w,z. The end result of the integration yields a formula for the vertical E-field in the central region of the conductor as

$$E_z = \eta \, (\pi\varepsilon_o)^{-1} \, [\text{Tan}^{-1}(w/2z) - \text{Tan}^{-1}(-w/2z)] \qquad (A1).$$

Close to the conductor surface, z is small and w >> z. In the limit that z is very small, the two arctan terms in Eq. (A1) become $\pi/2$ and $-\pi/2$ respectively, and combine to make $\pi$. Thus, in this small z limit, $E_z \sim \eta/\varepsilon_o$. For E-fields at locations where z > w, we can use the series expansion for the arctan functions to derive:

$$E_z \sim \eta w/\pi\varepsilon_o z \qquad (A2).$$

Note that the vertical E-field in the central region of the conductor (away from the edge) varies as 1/z, the distance from the conductor.

**Appendix 3: The Pressure Change near Ingenuity**

Ingenuity's mass is 1.8 kg. In order to obtain lift from the rotors, the lift force of the rotors on the gas has to exceed 6.7 N or a pressure across the blade plane of ~ 6.7 milliBar. The blades push gas fluid elements downward across blade plane forming a verticle downward-directed velocity field that extends approximately the scale size of the blades (~ 1-2 meters). Hence the local pressure gradient, dp/dx, near Ingenuity is ~ 3-6 milliBars/meter, which is a value that is about 1% of the ambient pressure – giving rise to a downward wind field that should be comparable or exceed those from dust devils (Ellehoj et al., 2010). These downward winds are expected to be incident with the surface. The surface will redirect the winds to move horizontally outward along the surface in a direction away from the rotorcraft. This effect is commonly called a 'downwash'. Surface grains would be lifted via saltation processes and forced upwards via the flow about local surface obstructions (mini-bumps or rocks). The analog is the lifted sand during landing from terrestrial helicopters. We also note that the outer edges of the blades possess a circular wind pattern (vortex) that would force grains upward at landing. See the diagram on http://www.copters.com/aero/hovering.html. While a more complete description would require a formal pressure and velocity field analysis, the fact that the pressure gradients are comparable or exceed those of dust devils imply a possible enhanced dust load at take-off and landing.

**Acknowledgements.** We gratefully acknowledge the NASA Internal Science Funding Model and its Fundamental Laboratory Research (FLaRe) program for supporting this research. Author AW also acknowledges NASA grant SSW-80NSSC17K0776 in supporting this work. We appreciate a discussion with Scott Guzewich regarding the atmospheric dust load. We also appreciate constructive discussions about the manuscript with the Ingenuity team members Havard Grip, Bob Balaram, and MiMi Aung.